\documentclass{article}
\usepackage{emulateapj,amsfonts,psfig}

\begin{document}
%\psfull

\title{The Brera Multi-scale Wavelet (BMW) ROSAT HRI source catalog \\
I: the algorithm}

\author{Davide Lazzati\altaffilmark{1}, Sergio Campana}

\affil{Osservatorio Astronomico di Brera, Via Bianchi 46, I-23807
Merate (Lc), Italy}

\author{Piero Rosati}

\affil{European Southern Observatory, Karl Schwarzschild Str. 2,
D-85748 Garching bei M\"unchen, Germany}

\and

\author{Maria Rosa Panzera and Gianpiero Tagliaferri}

\affil{Osservatorio Astronomico di Brera, Via Bianchi 46, I-23807
Merate (Lc), Italy}

\altaffiltext{1}{Dipartimento di Fisica, Universit\`a degli Studi di Milano,
Via Celoria 16, I-20133 Milano, Italy}

\submitted{Accepted for publication in ApJ, main journal}

\begin{abstract}

We present a new detection algorithm based on the wavelet transform 
for the analysis of high energy astronomical images.
The wavelet transform, 
due to its multi-scale structure, is suited for the optimal detection 
of point-like as well as extended sources, regardless of any loss of 
resolution with the off-axis angle.
Sources are detected as significant enhancements in the wavelet space,
after the subtraction of the non-flat components of the background. 
Detection thresholds are computed through Monte Carlo simulations in order 
to establish the expected number of spurious sources per field.
The source characterization is performed through a multi-source fitting in the 
wavelet space. The procedure is designed to correctly deal with very crowded
fields, allowing for the simultaneous characterization of nearby sources.
To obtain a fast and reliable estimate of the source parameters and 
related errors, we apply a novel decimation technique which, taking
into account the correlation properties of the wavelet transform, extracts
a subset of almost independent coefficients. 
We test the performance of this algorithm on synthetic fields,
analyzing with particular care the characterization of sources in
poor background situations, where the assumption of Gaussian statistics
does not hold. For these cases, where standard wavelet algorithms generally
provide underestimated errors, we infer errors through a procedure 
which relies on robust basic statistics.
Our algorithm is well suited for the analysis of images taken with the
new generation of X--ray instruments equipped with CCD technology which
will produce images with very low background and/or high source density.

\end{abstract}


\section{Introduction}

In the last 20 years the most widely used technique to analyze X--ray
images has been the so-called ``sliding box'' technique:  a box of
fixed size is passed across an image and sources are identified as
enhancements of the total photon counts in the box against the
background, which is independently estimated via local or global
measurements.
This technique has been  used in the past to build source catalogs
from imaging X-ray missions, such as Einstein (Harris 1990), 
EXOSAT (Giommi et al. 1991) and ROSAT/PSPC (White, Giommi \& Angelini 1994).
A different approach has been used in the compilation of the
ROSATSRC \markcite{zim94} (Zimmermann 1994) and the Rosat All Sky
Survey BSC \markcite{vog96} (Voges et al. 1996) catalogs, where maximum
likelihood filters were utilized for source detection and
characterization.

Although these catalogs have proven to be extremely useful, they also have 
shown important limitations. In deep confusion limited images,
especially crowded fields, or in the presence of very bright or extended
sources (such as clusters of galaxies and supernova remnants), these
techniques often tend to blend multiple sources into a single one,
to split a bright source in multiple detections and/or to misidentify small 
background fluctuations as real sources.
In addition, these catalogs provide a poor estimate (if any) of the
intrinsic source size and the distinction between extended and
point-like emission is hardly possible. This means that a visual check
``source by source'' is required in most cases in order to ``clean''
the source catalog.  An improvement of these techniques has been
presented by Vikhlinin et al. 1995, who substituted the sliding box
with a matched filter that reproduces at best the profile of point-like
sources. This method is optimal for detection of point-like sources,
but tends to miss faint sources that are spatially resolved by the
detector.
The next generation of X--ray missions, with very high
sensitivity and spatial resolution, will produce very deep images with
a very high surface density of X--ray sources (up to $1000/{\rm deg{^2}}$)
and a level of morphological detail comparable to that of optical images.
Therefore, it is important to develop new and more refined techniques for
source detection and characterization 
to fully exploit the scientific content of these vast datasets.

These problems and considerations led to the development of a 
new generation of detection algorithms in which the square window of 
the sliding box technique is replaced with a family of filters of 
different sizes with particular morphological 
and functional properties: the Wavelet Transform (hereafter WT).
The WT is a mathematical tool capable of decomposing an image in a set
of sub-images, each representing the image features at a given scale.
Sources with sizes ranging from the intrinsic instrumental resolution
up to a significant fraction of the whole image can be efficiently extracted 
at the WT scale that better matches the source size.
Moreover WT-based detection algorithms are insensitive to large scale
background fluctuations provided that the source size is different
from this scale, a typical situation in high energy images.

The application of WT-based algorithms to the automatic
detection and characterization of sources in X--ray images was first used
by Rosati et al.\markcite{ros93,ros95a,ros95b} 1993, 
1995 (see also Rosati 1995), who exploited the
multi-scale capabilities of this method to compile a catalog of 
extended sources in ROSAT-PSPC deep pointed observations. 
This work led to the compilation of the ROSAT Deep Cluster Survey
\markcite{ros95b,ros98} 
(RDCS; Rosati et al. 1995, 1998), a deep survey of galaxy
clusters over $\sim 50$ square degrees.
More recently, WT based techniques have been developed and applied by
other groups for the analysis of ROSAT, mainly PSPC, images 
\markcite{vik95,gre95,har96,pis97,dam97a,dam97b,pis98} (Grebenev et al. 1995; 
Harnden et al. 1996; Pislar et al. 1997; 
Damiani et al. 1997a,b; Vikhlinin et al. 1998), even if a complete catalog 
of WT detected sources has not been published to date.

In this and in the accompanying paper \markcite{cam98} (Campana et al. 
1999; Paper II hereafter) we present a multi-scale wavelet-based
algorithm which generalises and combines those used by 
Rosati et al. 1995 and Lazzati et al. 1998,
and its application to the complete 
archival data of the ROSAT High Resolution Imager (HRI). 
The paper is organized as follows: in 
Sect.~\ref{algosec}~and~\ref{charasec} we describe
how the detection and characterization algorithm works, while in 
Sect.~\ref{simulsec} we test its performances with simulations.
Finally, in Sect.~\ref{disc} we summarize our results and briefly 
discuss the applications of our algorithm.

\section{The detection algorithm}
\label{algosec}

\begin{figure}
\psfig{figure=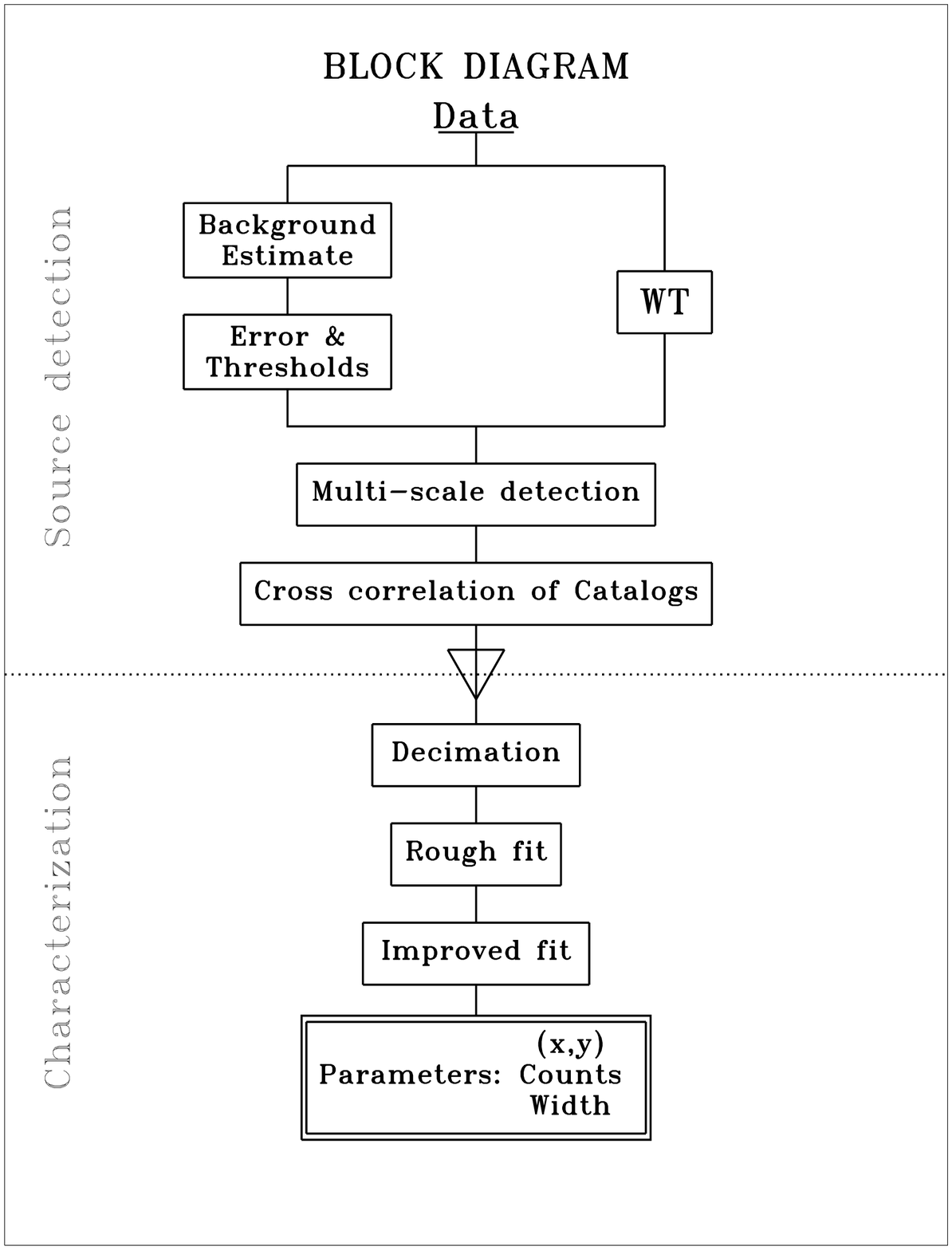}
\figcaption{{Block diagram of the detection and characterization algorithm.}
\label{blops}}
\end{figure}

The basic steps of the detection algorithm we developed
are summarized in Figure~\ref{blops}, while a more detailed description of
single steps is given in the following sections.

First, a wavelet convolution is performed on the X--ray image
and an average background value is estimated with a $\sigma$-clipping
algorithm (see Sect.~\ref{backsec}).
Given the measured background and referring to the simulations 
described in Sect.~\ref{simulsec}, a detection threshold is computed
constraining the total average number of spurious detections
in the whole image.
The source candidates are selected as significant (above the threshold) 
peaks in each independent wavelet sub-image and a complete catalog is 
obtained cross-correlating the single-scale catalogs to eliminate 
multiple-scale detections of the same source and the large-scale
blending of nearby small-scale sources.

At this stage, the catalog merely consists of a coarse position
and a rough estimate of the count rate and size of the sources.
The second part of our algorithm (see lower box of Figure~\ref{blops})
deals with the refinement of these quantities along with their
uncertainties and leads to the complete source characterization.
To this aim, a model source with a Gaussian point spread function
(hereafter PSF) is fitted in wavelet space
with a multi-dimensional procedure which takes into account 
the behavior of the wavelet coefficients both in space and scale. 
Up to 20 neighbor sources can be fitted
simultaneously in the deblending process, thus allowing  
the characterization of faint objects located near bright sources 
\markcite{laz98} (see also Lazzati et al. 1998).

As a final result, the algorithm produces - in a fully automated
way - a catalog of sources with position, count rate and apparent 
size along with corresponding uncertainties. For each source the associated
probability of spurious detection is also derived.

\subsection{The wavelet transform}

The wavelet transform  of a real function $f$ 
is the convolution product 
between $f$ and a class of analyzing wavelets 
$\psi_a(x,y)$, all derived by dilation from a single 
function $\psi_1(x,y)$, called mother 
wavelet:

\begin{equation}
\psi_a(x,\,y) = \frac 1a \psi_1(\frac xa,\, \frac ya).
\end{equation}

\noindent
Any function $g(x,y)$ can be used as the mother for a bidimensional
wavelet transform if it satisfies the following conditions:

\begin{eqnarray}
\int_{\mathbb{R}^2} |g(x,y)|^2 \, dx \, dy &<& \infty \nonumber \\
\int_{\mathbb{R}^2} g(x,y) \, dx \, dy &<& \infty 
\end{eqnarray}

The above properties assure the existence of the transform of any
square integrable function (as astronomical images are) and,
more importantly,
the fact that, whatever the function $f$ and the scale $a$
of the transform, the mean value of the transformed image is null.
This provide an objective and automatic background subtraction
in detection applications of the WT.

For a detailed and rigorous analytical description of
the wavelet operators, we refer the reader to the
extensive reviews published in the astronomical
literature \markcite{gre95,sle94} 
(e.g. Grebenev et al. 1995; Slezak et al. 1994)
or to more general discussions \markcite{mal92,you93}
(Mallat \& Hwang 1992; Young 1993)
and references therein.

In this work we adopt a discrete approximation of the wavelet operator
\markcite{gre95,laz98} (see, e.g. Grebenev et al. 1995; Vikhlinin et al. 1998; 
Lazzati et al. 1998),
which keeps all the properties described above provided that the integrals 
are replaced with discrete summations. This allows faster computations
and a reduction of the cross talks between adjacent pixels and
scales. The continuous approach, which has been adopted by - e.g. -
\markcite{dam97a,dam97b} Damiani 
et al. 1997a,b, allows the calculation of the
transform at whatever scale and position. In this scheme, 
source characterization proceeds through an iterative
refinement of the position and scale at which the transform
is calculated.  

Our WT computation is
based on the multi-resolution theory \markcite{far92} 
(Farge 1992) and the ``\`a trous'' 
algorithm \markcite{bij92} 
(Bijaoui \& Giudicelli 1992). We use the `mother' developed by
\markcite{sle95} 
Slezak et al. 1995, which can be analytically approximated with
the difference of two Gaussians:
\begin{eqnarray}
\psi_1(r) &=& A \, \exp \left[-{\frac{r^2}{2\sigma_A^2}}\right] -
B \, \exp\left[-{\frac{r^2}{2\sigma_B^2}}\right] \label{pareq} \\
A &\simeq& 0.597; \;\;\;\;\; \sigma_A \simeq 0.687 \nonumber \\
B &\simeq& 0.264; \;\;\;\;\; \sigma_B \simeq 1.030 \nonumber 
\end{eqnarray}

\noindent 
where the $(x,y)$ dependence has been turned into radial
due to the symmetry of the mother. This symmetry gives us
a new property of the WT operator: it is insensitive to
first order derivatives of the transformed function, i.e. the
transform of a plane is constantly equal to zero.
This means that smoothly varying backgrounds will not affect
the significance of the detections.
The multi-resolution technique provides us with a sampling on scales
logarithmically spaced by a factor of two. According to the wavelet theory 
(see e.g. \markcite{you93} Young 1993) 
this provides the best compromise between 
completeness and correlation (i.e. redundancies in the wavelet coefficients):
a finer sampling would cause an oversampling while a coarser sampling
would produce a loss of information. We hence have:

$$a_j = 2^j; \qquad j\in \mathbb{N}$$

\noindent
The sampling in the $x$ and $y$ is performed pixel by pixel,
at any scale. This produces an heavy 
oversampling of the transform, especially at the largest scales,
but it does not affect the detection. 
In the second part of the algorithm, when sources are characterized,
a decimation process is applied in order to minimize this
oversampling (see Sect.~\ref{decimasec}).

\subsection{Detection thresholds}
\label{threshsec}

Due to the structure of the wavelet operator, the statistics of
coefficients in the WT space cannot in general be handled analytically.
If, however, the function $f$ is a white Gaussian noise
with mean $\bar{d}$ and dispersion $\sigma_d$, we have that the 
individual WT coefficients are Gaussian distributed according to the law:

\begin{eqnarray}
p(WT) &=& \frac {1}{\sqrt{2\pi\sigma^2}} \, \hbox{exp} \Big (-\frac{WT^2}
{2\sigma^2} \Big ) \label{wtprob} \\
\sigma &=& \sqrt{\sigma_d^2 \; \int |\psi|^2}
\label{flatsig}
\end{eqnarray}

\noindent
Unfortunately, this is not the case
for high energy astronomical images that can be approximated by a 
white Poisson noise of constant mean (the background\footnote{Here we 
neglect the vignetting and the effects described in 
Paper II.}), on which sources are overlaid.
Since our goal is to assess the probability that the WT peaks
are random background realizations, we can neglect the statistics 
of the sources and concentrate on the transform of pure
background.
It can be shown (see, e.g. \markcite{gre95} Grebenev et al. 1995) 
that Equation~\ref{flatsig}
still holds for a Poisson noise, with 

$$\sigma = <WT^2> - <WT>^2 = <WT^2>; \quad \sigma_d = \sqrt{ \bar d}$$

\noindent
however, the probability distribution given by Equation~\ref{wtprob}
is not followed anymore. Moreover, even if we knew 
exactly the distribution of the single coefficients, the WT
non-orthogonality would prevent us from computing the completeness 
and contamination of the catalog.
Thus, to overcome these problems we must rely on Monte Carlo simulations.

\begin{figure}
\psfig{figure=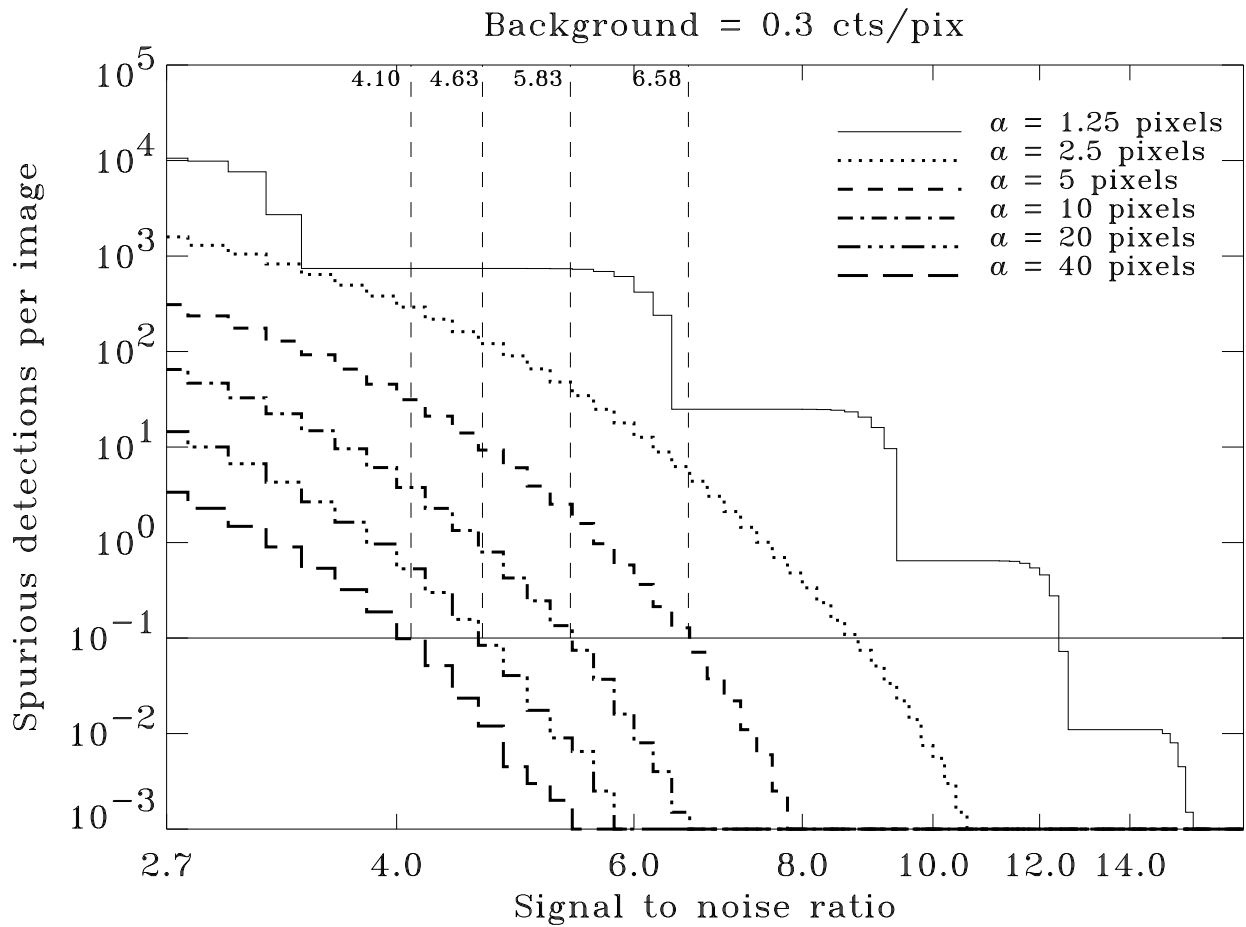}
\figcaption{{Threshold determination for detection of sources in wavelet space.
The plot shows the number of spurious detections with signal to noise larger
than a given value versus the signal to noise itself.
The six curves refer to six wavelet scales, from 1.25 to 40 pixels, for
simulations with a mean background of 0.3 counts per pixel.
The step-like behavior of the smallest scale curve is due to the
sensitivity of this scale on the discreteness of the background image.
For the four scales used in the search, the signal to noise thresholds
relative to a contamination of 0.1 sources per field have been marked.}
\label{soglieps}}
\end{figure}

For a set of mean background values between $10^{-4}$ and $10$ counts per 
pixel we simulated 5000 realizations of random fields, 
$512\times512$ pixels
each. Then we computed the WT and its standard
deviation with Equation~\ref{flatsig} and counted all local maxima
that exceeded a given signal-to-noise threshold (i.e. the ratio of the wavelet 
transform and the error given in Equation~\ref{flatsig}).
Since no source has been simulated, any peak in the transformed image
is spurious. Repeating this procedure with different thresholds
for each scale $a$ of the transform,
we have built the integrated probability function of finding
a random WT peak exceeding a given signal-to-noise.
Figure~\ref{soglieps} shows the typical output of a set of simulations
with a background of 0.3 counts per pixel. The number of 
WT peaks that exceeds a given signal-to-noise in the whole field is plotted 
versus the signal to noise itself 
for the different scales. To limit the number
of spurious detections to, e.g., one in ten fields, we draw a horizontal line 
corresponding to 0.1 sources and, scale by scale, we set the threshold
at the signal to noise where the horizontal line crosses
the corresponding curve.

Combining the results of all simulations we obtain the
integrated probability function $P$ which, given
the number of pixels of the search, the desired signal to noise threshold
and the mean value of the background counts, returns the mean expected number
of spurious detections. In the algorithm, this function must 
be inverted. In fact, when we run the WT algorithm on an image, we fix the 
expected contamination 
(the number of spurious detections in the whole image) and
we compute - as a function of the scale $a$ - the threshold
to be used.
Note that, since the number of independent 
points is proportional to the square of the inverse of the scale, 
the threshold applied on smaller scales will be 
higher than that on larger scales (see Figure~\ref{soglieps}).

In principle, the whole set of scales provided by the wavelet transform
can be used in the detection procedure. However, in the analysis
of high energy images - as those of the ROSAT HRI (see Paper II) -
we have a lower scale for the sources set by the instrumental resolution.
The analysis on very large scales, say half the size
of the image, is strongly contaminated by edge effects and sensitive to 
residuals from the image reduction. 
As a general rule, the best scales on which the analysis can be performed 
are the ones which closely match the instrumental PSF.
For these reasons, in the following of this paper and in the analysis of HRI 
images (Paper II), we will use only the following scales: 

\begin{equation}
a = 5; \, 10; \, 20; \, 40 \quad \hbox{pixels}
\label{scaleq}
\end{equation}

\noindent where the scale $a$ roughly measures the FWHM of the positive 
central part of the wavelet filter in pixels.

\subsection{Different scales cross-identification}
\label{crosec}

Since the source detection is performed on several scales and given the
non-orthogonality of the wavelet operator, a cross-correlation of
the single-scale catalogs must be performed.
This allows us to get rid of multiple identifications
of the same source at different scales and, more importantly, 
to discriminate true extended sources from a blending
of nearby point-like sources.
This latter distinction is not trivial,
since true extended sources often produce apparently significant peaks
in the lower scales, while nearby sources are inevitably blended
at larger scales.

First, we eliminate from the catalog multiple scale detection of
bright sources on the basis of position and signal-to-noise: among all sources
with compatible positions only the most significant (i.e. with the highest
signal-to-noise ratio) is held. Two sources detected at different wavelet 
scales but with nearby positions are matched in a single source 
if their distance is lower than the scale at which the most significant
has been detected.

A more difficult problem is dealing with the merging of nearby sources
at large scales. Two or more nearby point-like sources are usually detected 
as a single extended source at large scales. However, it is also
possible that a real extended feature is overlaid on them
(a typical situation in supernova remnants and in nearby clusters of 
galaxies). 
Our cross-correlation scheme held the large scale
detection as a true source only if its significance (signal-to-noise) is larger
than that of all lower-scale sources overlaid on it.
This scheme slightly favors point-like sources. 

The worse situation is provided by a group of weak and close sources,
which are detected only at large scales as a single extended feature. 
This is an intrinsic limitation of the detector resolution that cannot 
be solved by a detection 
algorithm, however refined (see, e.g. Damiani et al. 1997a).

Rosati et al. 1995 describe a different scheme, in which all sources are 
detected and characterized for each scale separately. A subsequent step 
consists in the cross-correlation of different scale catalogs to get rid of  
multiple and false detections by means of positional coincidence and fitted 
count-rates. Damiani et al. 1997a cross-identify only sources detected at 
consecutive scales and match sources if their distance is less than the scale 
size corresponding to the maximum signal-to-noise (or 1.5 times the detector 
PSF). 

Albeit simple, our scheme has been shown to be powerful and robust in the analysis of 
ROSAT HRI fields, that can be very crowded of point sources (see Paper II, in particular 
Figure 10).

\section{Source characterization}
\label{charasec}

After the whole procedure described in last section, we are left with a list of sources
with a rough determination of position (the center of the
pixel with higher WT coefficient), source size (the scale of the WT where the signal
to noise is maximized) and total number of photons 
(the value of the maximum WT coefficient).
Several schemes have been proposed in the literature to
refine these estimates in wavelet space; these can be divided in 
two main groups: fitting the multi-scale profile \markcite{gre95,ros95b}
(e.g. Grebenev et al. 1995;
Rosati et al. 1995) or searching for the best position and scale of the source 
recomputing the WT in a narrower grid both in space and in scale 
\markcite{dam97a,dam97b} (Damiani et al. 1997a,b).

Our approach is to fit source profiles both in space 
\markcite{ros95b} (as in Rosati et al. 1995) 
and in scale (as in \markcite{gre95} Grebenev et al. 1995). 
This allows us to refine simultaneously
the position, size and flux of the sources without the need
of weight summing parameters derived at different scales. 
The narrower grid method 
should in principle be more robust, since it does not involve
multi-parameter fitting procedures that could fail when run automatically. However, such a method needs analytic WT computations (instead of our
numeric method) and makes it difficult to deal with the strong cross
terms in WT space in presence of nearby sources and, more generally, in
crowded fields.

\subsection{Decimation technique}
\label{decimasec}

To apply standard routines ($\chi^2$ minimization)
to the source characterization in wavelet space, we must
first assure that the hypothesis of independence between 
input data is appropriate. As stated above, this is not the case, 
mostly at larger scales.
Although \markcite{ros95a} 
Rosati 1995 has shown that neglecting the correlations
between adjacent data does not significantly 
compromise the result of the fit,
the uncertainties on extracted parameters are highly underestimated
and the oversampling in $x,y,a$ lengthens considerably the computing time.

To obtain a fast, robust and unbiased estimate of the source parameters and 
related errors, we have developed a decimation technique that, taking
into account the correlation properties of the WT, extracts
a subset of quasi-independent coefficients from the whole transform (see also 
\markcite{laz98} Lazzati et al. 1998).

\begin{figure}
\psfig{figure=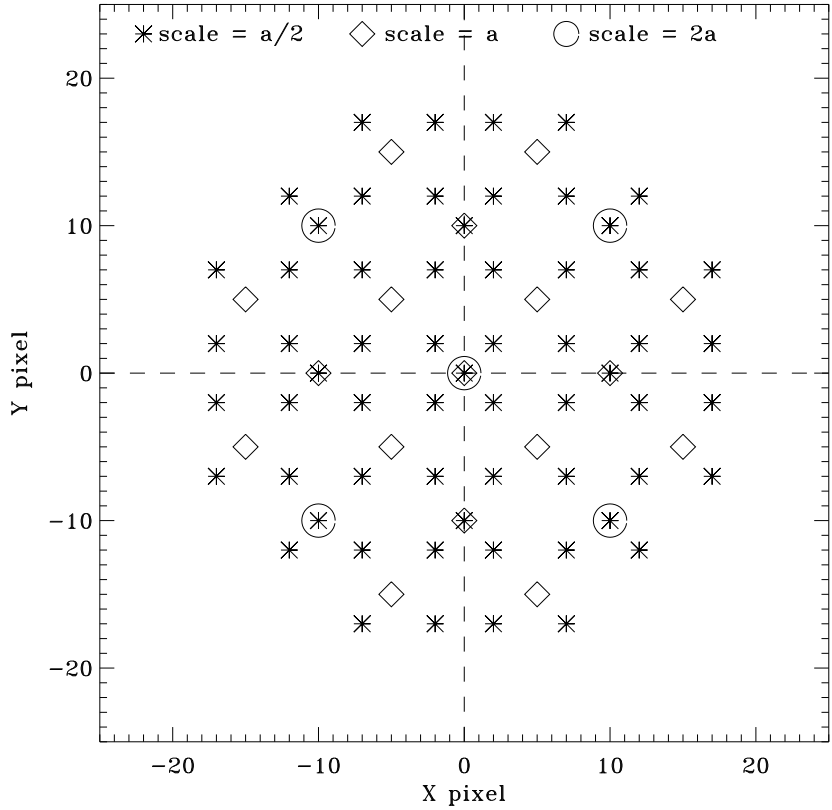}
\figcaption{{Decimation pattern for the three selected WT scales: 
asterisks mark the smallest scale ($a/2$); 
diamonds the median scale ($a$) and circles the largest scale ($2\,a$).
Note that the largest the scale, the most sparse the WT sampling.} 
\label{decima}}
\end{figure}

Before applying the decimation, we first select a subregion of the
transform around the source position.
For each detected source, named $\tilde a$ the most relevant scale of 
the detection, a region of radius  $4 \, \tilde a$
centered on the source position is extracted from the whole transform
and the scale range is reduced to the set 
$\{\tilde a/2,\, \tilde a,\, 2\, \tilde a\}$. 
Scales smaller than $\tilde a/2$ are dominated by noise, 
while those larger than $2 \, \tilde a$ suffer from the contamination 
from adjacent sources. 
To extract a subset of wavelet coefficients which is almost
independent, the minimum distance between two coefficients
at a given scale must be almost equal to the correlation length of the
transform at this scale, which is proportional to the scale itself. 
From the analysis of the autocorrelation function of the transform of
a pure white noise, we find that the correlation length is given by:
$c_l(a) \sim 1.7 \, a$ (see also \markcite{laz96} Lazzati 1996).
At the scales mentioned above we should have, in principle, 
69, 17 and 4 orthogonal wavelet coefficients, respectively.
Since the most significant coefficient at all scales is the coefficient
centered on the source position, we keep it in the decimated set.
We finally obtain 61, 17 and 5 coefficients for the three scales.
The position of this coefficients for the case $\tilde a = 5$
is plotted in Figure~\ref{decima}.
These extracted coefficients provide us with a small set of 
almost orthogonal\footnote{It is not possible to extract a set of rigorously 
orthogonal 
coefficients, but this decimation proves to be successful {\it a posteriori}
since standard fitting routines produce reliable results.} and 
Gaussian distributed points.

\subsection{Source fitting}

The model we fit to the decimated set of WT coefficients is the transform
of a bidimensional Gaussian. This transform can be computed analytically
and can be written as:

\begin{eqnarray}
\tilde f (x,y) &=& 2\pi I_0 \bigg [ A \sigma_A^2 \, G\Big(x,y,x',y',
\sqrt{\sigma^2 + 2^{2j}\sigma_A^2} \Big) - \nonumber \\
&-&  B \sigma_B^2 \, G\Big(x,y,x',y', \sqrt{\sigma^2 + 2^{2j}\sigma_B^2} 
\Big) \bigg] \label{treq}
\end{eqnarray}

\noindent
where $I_0$ is the integral of the source, $\sigma$ its width,
$A$, $B$, $\sigma_A$ and $\sigma_B$ are given in Equation~\ref{pareq}
and $G$ is defined as:

\begin{equation}
G(x,y,x',y',\sigma) = \frac1{2\pi\sigma^2}\exp \bigg[ 
-\frac{(x-x')^2+(y-y')^2}{2\sigma^2} \bigg]
\end{equation}

\noindent
The output of the procedure is hence a set of values
$\{x', y', I_0, \sigma \}$ along with their errors.

Since our fitting procedure involves scales larger than
the natural source scale, it is possible that the WT coefficients
are affected by the presence of nearby sources.
To deal with this problem, multiple source fitting (up to
20 simultaneous sources) can be performed. Thanks to the linearity
of the WT operator, the transform of $N$ sources in 
a small area is:

$$\tilde f (x,y) = \sum_{i=1}^N \tilde f_i (x,y)$$

\noindent
where $\tilde f_i (x,y)$ is given by Equation~\ref{treq}.
In practice, since it would be meaningless to fit $20 \times 4 = 80$
parameters with only 83 (i.e. $5+17+61$) data and since weak sources
could hardly affect the fit of a strong source, the procedure is structured
as follows. For each source, the algorithm searches for
all sources within ten times its estimated width.
In this set of sources, the parameters of those already fitted
are freezed while the total counts and sizes of all the other sources are free
in the fitting routine. 
When all sources have been fitted once,
the procedure is repeated (improved fit in Figure~\ref{blops}), 
but this time all nearby sources are held fixed.
This multi-source fitting allows a more precise determination of the source 
fluxes. This is particularly true in the case of very nearby 
sources where the influence of the negative wings of one source modifies 
the value of the other source transform (see e.g. 
\markcite{laz98} Lazzati et al. 1998).

\subsubsection{Error determination}

If the number of background and source counts is sufficient 
($\gtrsim 5\times 10^{-2}$~counts per pixel),
a reliable estimate of the parameter uncertainties can be directly 
obtained from the covariance matrix (see Sect.~\ref{simulsec} for more details).
However, a more detailed computation of the error on WT coefficients is needed. 
In fact, Equations~\ref{wtprob} and~\ref{flatsig} are no longer valid to estimate 
the WT fluctuations in presence of sources (i.e. non-flat components).
A more general form for Equation~\ref{flatsig} is given in \markcite{gre95} 
Grebenev et al. 1995:

\begin{equation}
\sigma(x,y) = \sqrt{
f(x,y) \ast |\psi|^2}
\label{nfsig}
\end{equation}

\noindent
where $\ast$ represents the convolution product in $\mathbb{R}^2$
and $f(x,y)$ is the function to be transformed. Poisson statistics for $f$ 
has been assumed. Note that Equation~\ref{nfsig} is consistent with
the results given above when $f=\hbox{const.}$
(cf. Equations~\ref{wtprob},~\ref{flatsig}).

When dealing with very low backgrounds ($\lesssim 5\times 10^{-2}$~counts per pixel), 
in particular at the lowest scales, the error determination suffers from the strong 
departure of the WT statistics from the Gaussian case. As a result, errors
are generally underestimated.
In this case a precise determination in WT space would imply
a fitting procedure based on WT statistics which, as stated above,
cannot be handled analytically.
However, from basic statistics, we can obtain reliable estimates
of errors with  an {\it a priori} method.
Let's consider position $x'$ and $y'$. 
If we have a Gaussian shaped 
spatial distribution
of $N$ photons with a width $\Delta$, the precision on the
determination of the centroid is limited by:

\begin{equation}
\sigma_{x'} \simeq \sigma_{y'} \simeq \frac{\Delta}{\sqrt{N-1}}
\label{xunc}
\end{equation}

\noindent
while the error on total number of counts will be given by Poisson statistics:

\begin{equation}
\sigma_{N} \simeq \sqrt{N}
\label{func}
\end{equation}

\noindent
The intrinsic limit on the width estimate can be expressed through the 
rms scatter of photon positions:

\begin{equation}
\sigma_\Delta \simeq 0.5 \, \frac{\Delta}{\sqrt{N-1}}
\label{sunc}
\end{equation}

The final error on fitted parameters is set to be the maximum value between 
that estimated from the covariance matrix and the above equations.

\section{Testing the algorithm}
\label{simulsec}

To test the reliability of the source characterization and of the
parameter uncertainties, we have simulated a set
of $21$ fields with $64$ sources each, for a set of backgrounds
ranging from $10^{-3}$ to $10^{-1}$~counts per pixel. Hence, 
for each background value we have 1344 synthetic sources.
The source positions are fixed to an equally spaced 
$8\times8$ grid in order to test the
accuracy of the fit without edge problems and source confusion
(see also Paper II for a complete discussion of this problem).
To reproduce at best an astronomical image, 
the total counts of sources have been simulated according to an
Euclidean ${\rm Log}\, N - {\rm Log}\, S$, while all sources have been assumed 
point-like with a FWHM~$\simeq 6$~pixels.

We have then applied the full procedure described above, including background
determination and source detection with a threshold of 0.1 spurious sources 
per simulated field.

\subsection{Background determination}
\label{backsec}

The estimate of the mean background value 
is carried out with a $\sigma$-clipping method.
This procedure relies on the hypothesis that the statistics of the
image is dominated by two independent distributions: the background counts
and the source counts. If the average values of these distributions are 
strongly different (which is very often the case when dealing with
 high energy astronomical images) the contribution of sources 
can be iteratively erased and a robust mean background value measured.

To reach this goal the mean and standard 
deviation of the whole image are calculated and pixels 
with a number of counts that exceeds the mean value 
by more than 3 standard deviations are flagged.
The process is repeated iteratively on unflagged pixels until the mean and the
standard deviation values converge. 
Since the $3 \, \sigma$-clipping implies  Gaussian statistics,
the images are binned to obtain an initial mean number of counts
per pixel of at least 10.

\subsection{Results}
\label{risultati}

These simulations have revealed that as long as the distribution 
of WT coefficients is well approximated by a Gaussian function, 
the procedure works very well.
With small background values - below $\sim 5\times 10^{-2}$ counts per pixel -
the distribution  in the lowest scales of WT space becomes 
increasingly Poissonian  and the covariance matrix gives underestimated 
errors, even if the parameter evaluation remains reliable.
In this case, the uncertainties given by
Equations~\ref{xunc},~\ref{func}~and~\ref{sunc} turn out to be more
accurate than those obtained in the fitting procedure.
Sources with even five input photons (the minimum input value) are detected in
the lowest background simulations. This is not a surprising result because,
as pointed out by Damiani et al. 1997a, in principle it is possible to detect 
sources with 3 photons only.

\begin{figure}
\psfig{figure=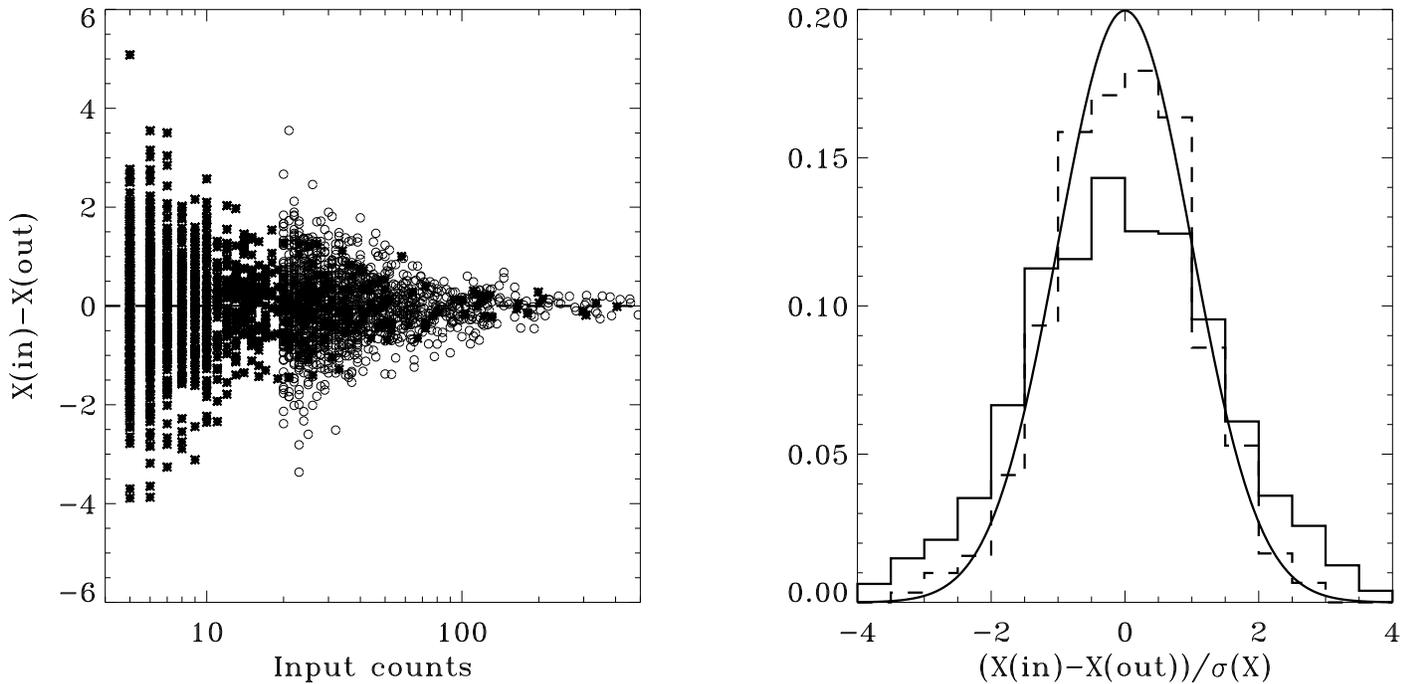}
\figcaption{{Simulation results on position accuracy. 
Left panel shows the absolute error in the X-axis position
determination in pixels versus the input counts of the simulated source.
Asterisks refer to the $10^{-3}$~counts per pixel background simulations
while circles to the $10^{-1}$~counts per pixel simulations.
Right panel shows the distribution of position offset normalized
to their estimated errors. The curve is a normal Gaussian. 
Dashed and solid lines refer to the lower and higher 
backgrounds, respectively.
}\label{s1ps}}
\end{figure}

\begin{figure}
\psfig{figure=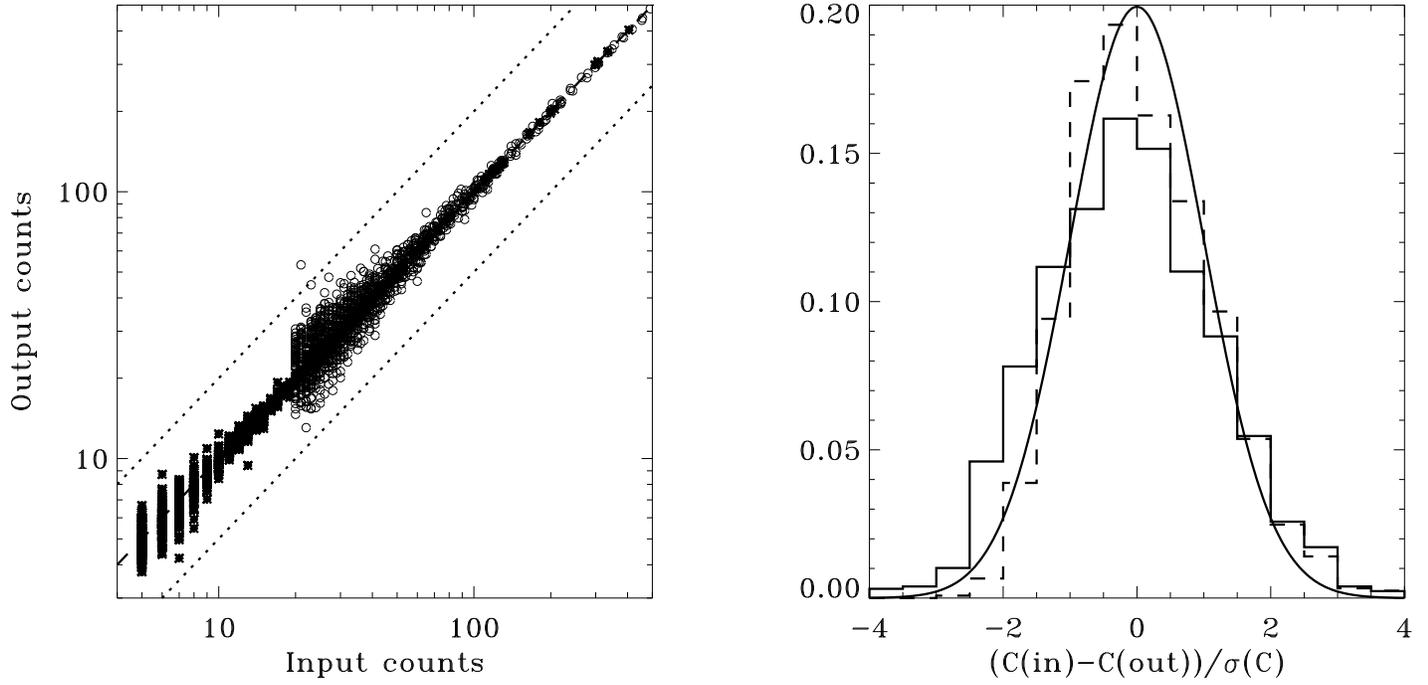}
\figcaption{{Simulation results on total counts determination accuracy. 
Left panel shows the output counts
versus the input counts of all simulated sources.
Plotting symbols are analogous to those of Figure~\protect\ref{s1ps}.
Right panel shows the distribution of count differences normalized
to their estimated errors. The curve is a normal Gaussian. 
Dashed and solid lines refer to the lower and higher 
backgrounds, respectively.
}\label{s2ps}}
\end{figure}

\begin{figure}
\psfig{figure=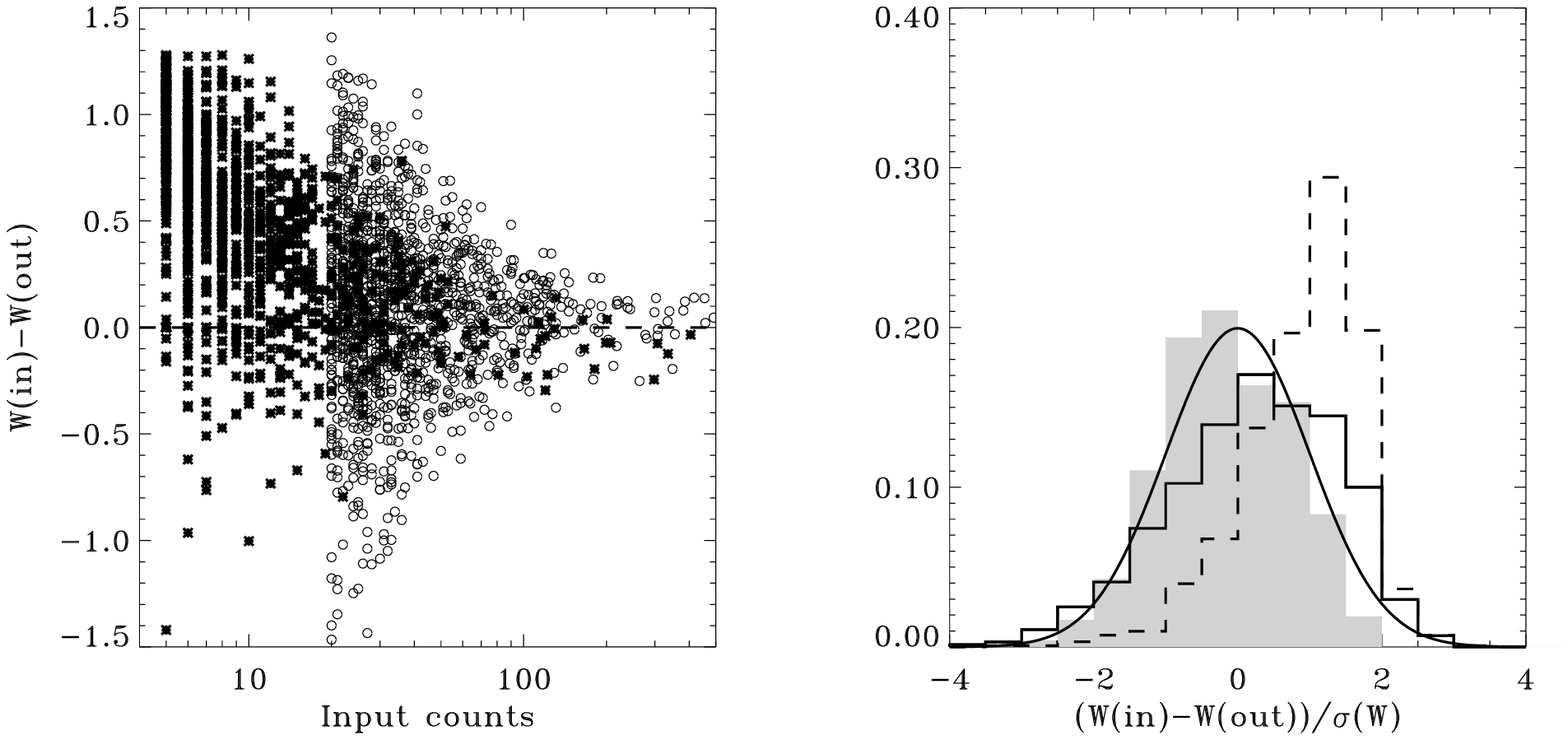}
\figcaption{{Simulation results on source width accuracy.
Left panel shows the absolute error in the width determination
versus the input counts of the simulated source.
Plotting symbols are analogous to those of Figure~\protect\ref{s1ps}.
Right panel shows the distribution of size determination differences normalized
to their estimated errors. The curve is a normal Gaussian.
Dashed and solid lines refer to the lower and higher 
backgrounds, respectively.
The shaded region shows the histogram for the lower background when only
bright sources ($>100$~counts) are considered.
}\label{s3ps}}
\end{figure}

Figures~\ref{s1ps},~\ref{s2ps}~and~\ref{s3ps} show the result of these
simulations. In all figures, the first panel shows the absolute
discrepancy between the input and output parameters (position, total
counts and width, respectively) versus the input counts. In the right
panels the  distribution of the absolute discrepancy between the input
and output parameters, divided by the their estimated errors, are
compared with a normal Gaussian distribution.  Asterisks and dashed
lines refer to the lower background ($10^{-3}$~counts per pixel)
simulations, while circles and solid lines to a higher background case
($10^{-1}$~counts per pixel). 

For low background simulations (dashed lines in
Figures~\ref{s1ps},~\ref{s2ps}~and~\ref{s3ps})
the correlation matrix gives underestimated errors and Equation~\ref{xunc}, 
\ref{func} and \ref{sunc} are always used. 
A Kolmogorov-Smirnov (KS) test on the whole set of
simulated sources confirms the visual impression that dashed histograms
are not incompatible with the Gaussian solid line in the figures.
For high background simulations (solid histograms)
the correlation matrix is commonly used.
Considering the whole set of simulated sources, errors are slightly
underestimated (20\%, especially in Figure~\ref{s1ps}). 
However, a closer inspection reveals that when only bright ($>50$~cts) 
sources are selected, the estimated 
errors are Gaussian distributed.
We note that the overestimation of fluxes near the detection threshold is 
common, due to the fact that sources on top of positive background fluctuations are 
preferentially selected.

Figure~\ref{s1ps} shows that source positions are determined
with an accuracy which is of the order of the source half width in the low 
signal-to-noise regime.
A similar result is obtained for the total
counts (Figure~\ref{s2ps}). Only for one source out of $2500$
the total counts evaluation has failed by more than a factor of two.
This faint source, whose flux and size measurements are affected by a large error, 
lays in the close vicinity of one of the four significant positive background 
fluctuations expected (40 simulated fields with 0.1 spurious sources each).
The algorithm merged the true source and the fluctuation in a higher flux extended 
feature. For what concerns the measurement of the source size, Figure~\ref{s3ps} shows
that particularly in the lower background simulations the width of the source
is underestimated. This is due to the fact
that, since the vast majority of sources in our simulations are at the limit
of detection, those that randomly happen to be more
compact have a higher signal-to-noise ratio and are hence 
preferentially selected by the algorithm against those with lower surface
brightness.
This bias is present only for weak sources in very low background
images and  would cause the misidentification of extended sources as
point-like and not vice-versa.  The shaded region in the right
panel of Figure~\ref{s3ps} shows how this bias is considerably reduced
when only sources with more than 100 counts are considered.

\section{Conclusions}
\label{disc}

We have presented a new detection algorithm based on the wavelet transform
for a multi-scale analysis of astronomical X--ray images which is suited 
for the detection and the characterization of both point-like and extended 
sources.
Given an image in the X-ray band, the algorithm produces a
catalog of sources characterized by position, count rate and size
along with associated errors.

The main difference of our technique with respect to other WT-based 
algorithms lies in the source characterization.
Our algorithm uses, for the first time, a 
fitting procedure that models the source in WT-space
by taking into account the spatial and multi-scale behavior 
simultaneously. Moreover, this procedure is performed on a decimated set of
wavelet coefficients to drastically reduce the cross-talk between adjacent
pixels and scales.

Future high energy missions
equipped with CCD detectors, such as XMM, AXAF and JET-X, will combine
a high effective area with a small instrumental background. In these
cases, current wavelet implementations which rely on Gaussian
statistics for the source characterization may not lead to optimal results.
The problem of crowded fields has been dealt with a multi-source characterization 
procedure, where up to 20 sources are fitted simultaneously. 
In very low background images, this fitting procedure produces reliable 
parameter estimates but underestimated errors. 
In these cases we have developed a different error evaluation procedure
which relies on basic statistics and which proves to be robust even with 
vanishing backgrounds.

The use of wavelet-based detection algorithms for the data analysis of
the new generation of X--ray missions will provide an accurate, fast
and user friendly source detection software.  A first application of
this algorithm to high resolution X-ray images obtained with the ROSAT
HRI is presented in the accompanying paper \markcite{cam98} (Campana et
al.  1999) and the analysis of the full set of archival HRI pointings
is underway for the construction of a complete wavelet-based catalog of
ROSAT HRI X-ray sources.

\acknowledgements{This work has been supported through ASI and CNAA grants.}

\newpage

\begin{references}

\reference{bij91} Bijaoui, A., \& Giudicelli, M. 1991, 
Experimental Astronomy, 1, 347

\reference{cam98} Campana, S., Lazzati, D., Panzera, M. R., 
\& Tagliaferri, G. 1999, ApJ in press (Paper II)

\reference{dam97a} Damiani, F., Maggio, A., Micela, G., \& Sciortino, S.
1997a, ApJ, 483, 350

\reference{dam97b} Damiani, F., Maggio, A., Micela, G., \& Sciortino, S. 
1997b, ApJ, 483, 370

\reference{far92} Farge, M. 1992, Annual Rev. of Fluid Mech., 24, 395

\reference{gio91} Giommi, P., Tagliaferri, G., Beuermann, K., et al. 
1991, ApJ, 378, 77

\reference{gre95} Grebenev, S. A., Forman, W., Jones, C., \& Murray, S. 
1995, ApJ, 445, 607

\reference{har96} Harnden, F. R, Caillault, J. P., Damiani, F., et al. 1996, 
in ``R\"ontgenstrahlung from the Universe'', Eds. Zimmermann, H.U., 
Tr\"umper, J., \& Yorke, H, p. 43

\reference{har90} Harris, D. E. 1990, ``The Einstein Observatory Catalog 
of IPC X-ray Sources'', Cambridge

\reference{laz96} Lazzati, D. 1996, 
Tesi di Laurea, University of Milano--Como

\reference{laz98} Lazzati, D., Campana, S., Rosati, P., Chincarini, G.,
\& Giacconi, R. 1998, A\&A, 331, 41

\reference{mal92} Mallat, S., \& Hwang, L. H. 1992, 
IEEE Trans. on Information Theory, 38

\reference{pis97} Pislar, V., Durret, F., Gerbal, D., Lima Neto, G. B.,
\& Slezak, E. 1997, A\&A, 322, 53

\reference{ros93} Rosati, P., Burg, R., \& Giacconi, R. 1993,
AIP Conference Proceedings, 313, 260, eds. Schlegel, E. M., \& Petre, M.

\reference{ros95a} Rosati, P. 1995, PhD. Thesis. Rome University

\reference{ros95b} Rosati, P., Della Ceca, R., Burg, R., Norman, C.,
\& Giacconi, R. 1995, ApJ, 445, L11

\reference{ros98} Rosati, P., Della Ceca, R., Norman, C.,
\& Giacconi, R. 1998, ApJ, 492, L21

\reference{sle94} Slezak, E., Durret, F., \& Gerbal, D. 1994, AJ, 108, 1996

\reference{vik95} Vikhlinin, A., Forman, W., Jones, C., \& Murray, S. 
1995, ApJ, 451, 542

\reference{vik98} Vikhlinin, A., McNamara, B. R., Forman, W., et al. 
1998, ApJ, 502, 558

\reference{vog96} Voges, W, Aschenbach, B., Boller, T., et al. 1996, 
IAU Circ., 6420

\reference{whi94} White, N. E., Giommi, P., \& Angelini, L. 1994, 
IAU Circ. 6100

\reference{you93} Young, R. K. 1993, 
``Wavelet theory and its applications'', Kluwer Academic 
Publishers

\reference{zim94} Zimmermann, H. U. 1994, IAU Circ., 6102

\end{references}
\end{document}